\documentclass[prl,aps,twocolumn,superscriptaddress,10]{revtex4-2}
\usepackage[english]{babel}
\usepackage{amsmath,amssymb,amsfonts,bbold}

\usepackage{graphicx}
\usepackage[colorlinks=True,linkcolor=blue,citecolor=blue,urlcolor=blue]{hyperref}
\usepackage{import}
\usepackage{booktabs}

\usepackage[dvipsnames]{xcolor}

\makeatletter
\renewcommand*{\fnum@figure}{{\normalfont\bfseries \figurename~\thefigure}}
\renewcommand*{\fnum@table}{{\normalfont\bfseries \tablename~\thetable}}

\newcommand{\ii}{\text{i}}

\makeatother

\begin{document}
\title{Non-Equilibrium Ising-Bloch Transition in Polarization Cavity Solitons}
\author{Juan Diego Mazo-Vásquez}
\email{juan-diego.mazo-vasquez@mpl.mpg.de}
\affiliation{Max Planck Institute for the Science of Light, Staudtstraße 2, 91058 Erlangen, Germany}
\affiliation{Department of Physics, Friedrich-Alexander Universität Erlangen-Nürnberg, Staudtstraße 7, 91058 Erlangen, Germany}
\author{Lewis Hill}
\affiliation{Max Planck Institute for the Science of Light, Staudtstraße 2, 91058 Erlangen, Germany}
\author{Flore K. Kunst}
\affiliation{Max Planck Institute for the Science of Light, Staudtstraße 2, 91058 Erlangen, Germany}
\affiliation{Department of Physics, Friedrich-Alexander Universität Erlangen-Nürnberg, Staudtstraße 7, 91058 Erlangen, Germany}
\author{Gian-Luca Oppo}
\affiliation{SUPA and Department of Physics, University of Strathclyde, 107 Rottenrow, Glasgow G4 0NG, Scotland, United Kingdom}
\email{g.l.oppo@strath.ac.uk}

\begin{abstract}
Polarization cavity solitons in coherently driven Kerr resonators undergo a sequence of dynamical bifurcations as the cavity detuning is increased. 
Here we show that a stationary localized state containing an Ising wall first becomes a fixed breather and then develops finite chirality and drift motion through a spontaneous non-equilibrium Ising–Bloch transition. 
Stability analyses identify the modes responsible for the breathing and the motion, revealing controllable dynamics of polarization cavity solitons.
\end{abstract} 

\maketitle

\emph{Introduction}--- Cavity solitons (CSs), a special kind of localized dissipative structures, emerge in driven nonlinear resonators through a balance between energy injection, dissipation, nonlinearity, and spatial coupling~\cite{Firth98, Ankiewicz2008, LugiatoBook15}. 
Their relevance and importance has recently been highlighted by applications in the generation of optical frequency combs~\cite{Pasquazi18, Oppo24}. 
Among localized dissipative structures, domain walls connect coexisting homogeneous states related by a discrete symmetry~\cite{HaeltermanPRE1994}, and can also arise in conservative multicomponent systems, such as the relative-phase walls of coherently coupled two-component Bose--Einstein
condensates~\cite{SonPRA2002}.
In parametrically driven systems, such domain walls may undergo a non-equilibrium Ising–Bloch transition (NIBT), in which a stationary Ising wall loses reflection symmetry and bifurcates into two Bloch walls of opposite chirality that propagate in opposite directions~\hbox{\cite{CoulletPRL1990, MichaelisPRE2001, deValcarcel2002, Clerc2005}}.
This transition has been observed experimentally in several nonlinear optical systems~\cite{Nasuno1995, EstebanMartin2005} and is generally associated with the interaction between a parity-breaking internal mode and the translational Goldstone mode of the wall~\cite{MichaelisPRE2001}.

Kerr resonators provide a rich platform for studying CSs. 
The dynamics of two orthogonally polarized intracavity fields are described by a pair of coupled Lugiato–Lefever equations (LLEs)~\cite{LugiatoPRL1987, Haelterman92}, whose cross-phase modulation enables spontaneous polarization-symmetry breaking~\cite{Geddes94, Woodley18}.
Recent work has revealed a new class of vectorial CS in normally dispersive Kerr resonators, termed polarization faticons~\cite{LucasPRL2025}. 
These structures consist of two bright temporal lobes with opposite polarization handedness, connected by an Ising domain wall in one of the linear polarization components. 
Faticons combine fast-time localization with polarization symmetry breaking and exhibit a singular structure~\cite{LucasPRL2025}.

The chirality originally associated with a faticon refers to the opposite polarization handedness of its two lobes. 
This is distinguished from the chirality of a Bloch wall, which characterizes the direction of phase rotation across the wall core and determines its direction of propagation.
Whether a polarization faticon can acquire this second form of chirality and undergo a spontaneous transition to a drifting state is a nontrivial question.
This is because stationary faticons lose their stability through a Hopf bifurcation leading to a time-periodic localized state~\cite{LucasPRL2025}.

Here we show that, as the cavity detuning increases, a stationary faticon first becomes a breather with a fixed Ising core through a Hopf bifurcation, and then undergoes an NIBT into a pair of drifting breathers with opposite velocities and domain-wall chiralities.
The critical Floquet eigenmode breaks the spatial parity of the faticon, while the drift velocity and chirality emerge continuously at the transition.
This extends the NIBT~\cite{GomilaPRL2015} to breathing vectorial localized of the Ising type structures in coherently driven Kerr resonators.

\begin{figure}[hbt!]
    \centering
    \includegraphics[width=\linewidth]{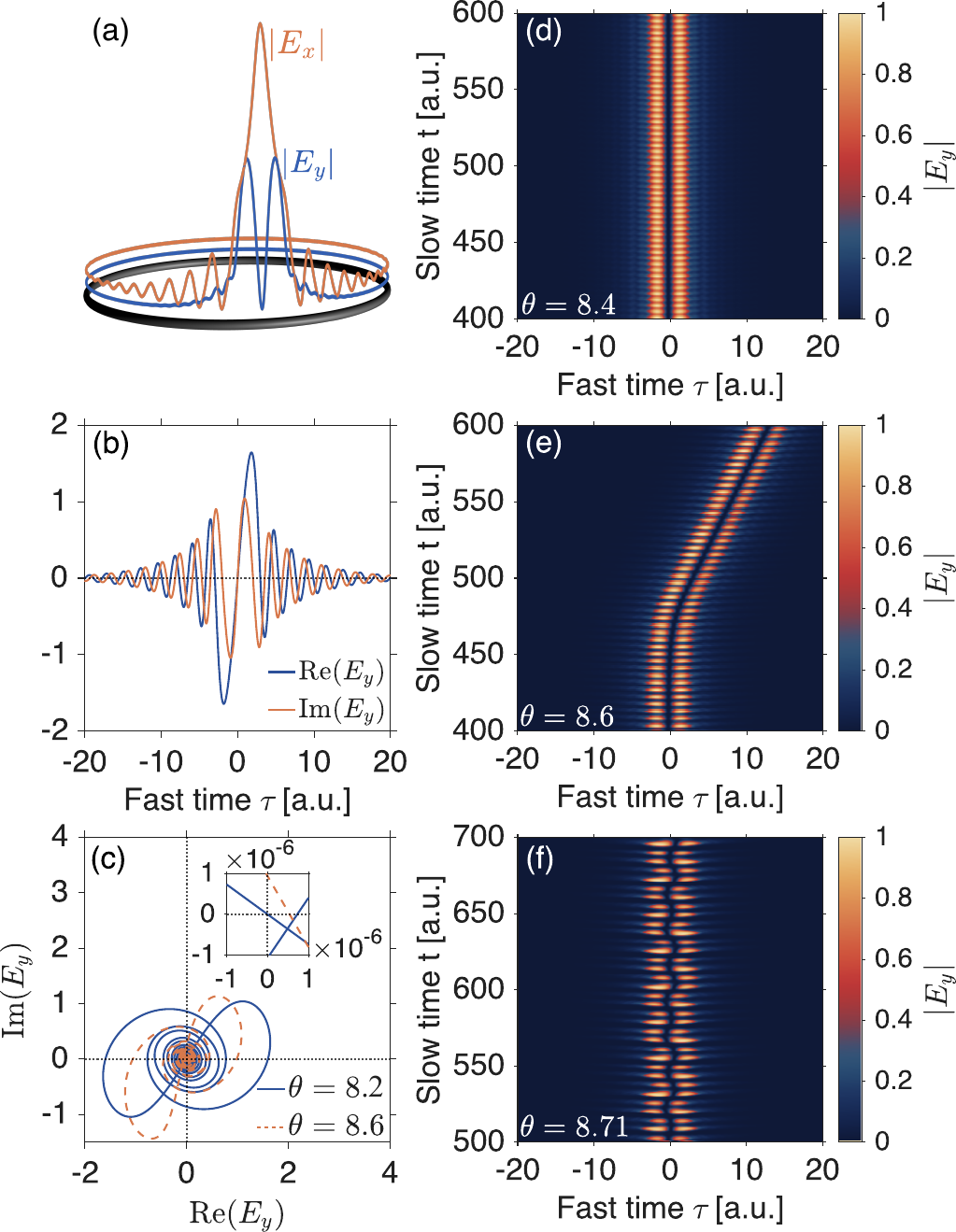}
    \caption{(a) Schematics of the amplitude profiles of the linearly polarized field components $E_{x,y}$ for a stationary faticon ($\theta=8.2$) propagating in a ring resonator. 
    (b) Real and imaginary parts of $E_y$ for the stationary solution.
    Both components vanish simultaneously at $\tau = 0$, identifying the singularity associated with the Ising-type PDW. 
    (c) Parametric representation of $E_y(\tau)$ in the complex plane before ($\theta = 8.2$, solid blue line passing through the origin) and after ($\theta = 8.6$, dashed orange line) the NIBT.
    (d–f) Slow-time evolution of $|E_y|$ for increasing cavity detuning: breathing Ising faticon ($\theta = 8.4$), drifting breathing faticon ($\theta = 8.6$), chaotic faticon with zero mean velocity ($\theta = 8.71$).
    All simulations were performed with $B=2$ and $X=16.5$. In (d)–(f), $|E_y|$ is shown in normalized units.}
    \label{fig1}
\end{figure}

\emph{Model}--- We consider the coupled LLEs describing the evolution of the two circularly polarized intracavity field envelopes~\cite{Xu2021, LucasPRL2025},
\begin{eqnarray}\label{eq:LLE}
\frac{\partial E_{\pm}}{\partial t} &=&  \Big[-\ii\eta\frac{\partial^2}{\partial \tau^2} + \ii\left(|E_{\pm}|^2 + B|E_{\mp}|^2 \right)-1-\ii\theta\Big]E_{\pm} \nonumber \\
&+& \sqrt{X/2}
\end{eqnarray}
where $t$ and $\tau$ denote the slow and fast times, respectively, $\theta$ is the normalized cavity detuning, $X$ is the normalized driving power, and $B$ is the cross-phase modulation coefficient.
Throughout this work we consider an isotropic Kerr cavity with normal dispersion ($\eta = 1$), for which polarization faticons were recently predicted and experimentally observed~\cite{LucasPRL2025}.
To characterize the internal structure of these localized states, we use the linearly polarized field components
\hbox{$E_x = (E_+ + E_-)/\sqrt{2}$} and \hbox{$E_y = (E_+-E_-)/(\ii \sqrt{2})$} (the corresponding equations are given in~\cite{SI}). 
The NIBT is encoded in the internal structure of the polarization domain wall (PDW) of the $E_y$ component, whose amplitude vanishes at the wall center (Ising state) and corresponds to a nested dark soliton.
Consequently, the analysis below focuses on the dynamics of $E_y$.
Equation ~\eqref{eq:LLE} exhibits two discrete symmetries: $(i)$ parity \hbox{$\mathcal{P}: \tau \rightarrow -\tau$}, and $(ii)$ fields exchange \hbox{$\mathcal{Z}: (E_+, E_-) \rightarrow (E_-, E_+)$}, or alternatively \hbox{$\mathcal{Z}: (E_x, E_y) \rightarrow (E_x, -E_y)$}.

\emph{Results}--- We begin by characterizing the internal structure of the polarization faticon and then analyze its dynamical evolution. 
Figure~\ref{fig1}(a) shows the two linearly polarized field components $(E_x, E_y)$ obtained from the stationary solution of Eq.~\eqref{eq:LLE} for a faticon. The $x$-polarized component $E_x$ forms a bright localized soliton, whereas the orthogonal component $E_y$ exhibits a dark localized structure composed of two symmetric lobes separated by a narrow central core of zero intensity. 
Figure~\ref{fig1}(b) displays its real and imaginary parts as functions of the fast time $\tau$. Both components vanish simultaneously at the center of the localized structure, where the field amplitude is zero while the optical phase experiences a discontinuous jump of $\pi$. 
Consequently, the central core constitutes an Ising PDW connecting the two lobes of opposite handedness and where $E_y = 0$ at the wall center while preserving reflection symmetry about it. 
A complementary representation is provided in Fig.~\ref{fig1}(c), where the trajectory of $E_y(\tau)$ is shown in the complex plane. 
Before the NIBT, for $\theta=8.2$, the curve passes through the origin, reflecting the simultaneous vanishing of the real and imaginary parts of the field at the wall center.
After the NIBT the $E_y(\tau)$ trajectory no longer intersects the origin and displays a continuous rotation in the complex plane (Fig.~\ref{fig1}(c)). 
This geometric feature is a signature of the transition from an Ising- to a Bloch-type domain wall and provides a direct visualization of the symmetry breaking underlying the onset of motion~\cite{CoulletPRL1990}.

Having established the internal structure of the localized state, we now examine its slow-time dynamics.
The remaining panels of Fig.~\ref{fig1} illustrate the evolution of the amplitude of $E_y$ for increasing cavity detuning while the driving power $X$ is kept fixed.
For detunings below the first instability threshold the faticon rapidly converges toward a stationary localized state after a transient. 
The wall remains at a fixed position at all times and no drift is observed, consistent with the Ising character identified in Figs.~\ref{fig1}(b,c).
In contrast to previously studied optical Ising walls~\cite{MichaelisPRE2001, SanchezMorcillo2005, EstebanMartin2005}, this wall is nested inside a bright soliton whose tails reach the same homogeneous steady state (HSS) on both sides, so that a single faticon is compatible with periodic boundary conditions.

Increasing the cavity detuning beyond the first threshold produces a qualitative change in the dynamics.
As shown in Fig.~\ref{fig1}(d), the localized structure undergoes periodic oscillations while remaining spatially fixed. The breathing develops through a Hopf bifurcation that destabilizes the stationary faticon without modifying the symmetry of its core.
Throughout the oscillation the domain wall remains at the same location, indicating that this periodic state still retains an Ising-type structure, here identified as a breathing Ising wall.

A second transition occurs upon further increasing the detuning. Figure~\ref{fig1}(e) shows that the breathing state eventually loses reflection symmetry and begins to drift with an average velocity $v$. The localized structure continues to breathe while translating through the resonator, demonstrating that the symmetry breaking acts on the periodic orbit generated by the Hopf bifurcation. 
This behavior is also revealed in a reference frame moving at the measured velocity $v$, defined through the coordinate $\tau'=\tau-vt$. In this frame, the position of the localized structure remains fixed, whereas its profile continues to oscillate periodically, see End Matter. 
The drifting and breathing faticon is associated with a dynamical NIBT, in which we move from an oscillating Ising wall to an oscillating and moving Bloch wall, a feature with analogies and differences with that described in~\cite{GomilaPRL2015} where the state before the transition to motion is an oscillatory state containing Bloch walls. 

\begin{figure}
    \centering
    \includegraphics[width=\linewidth]{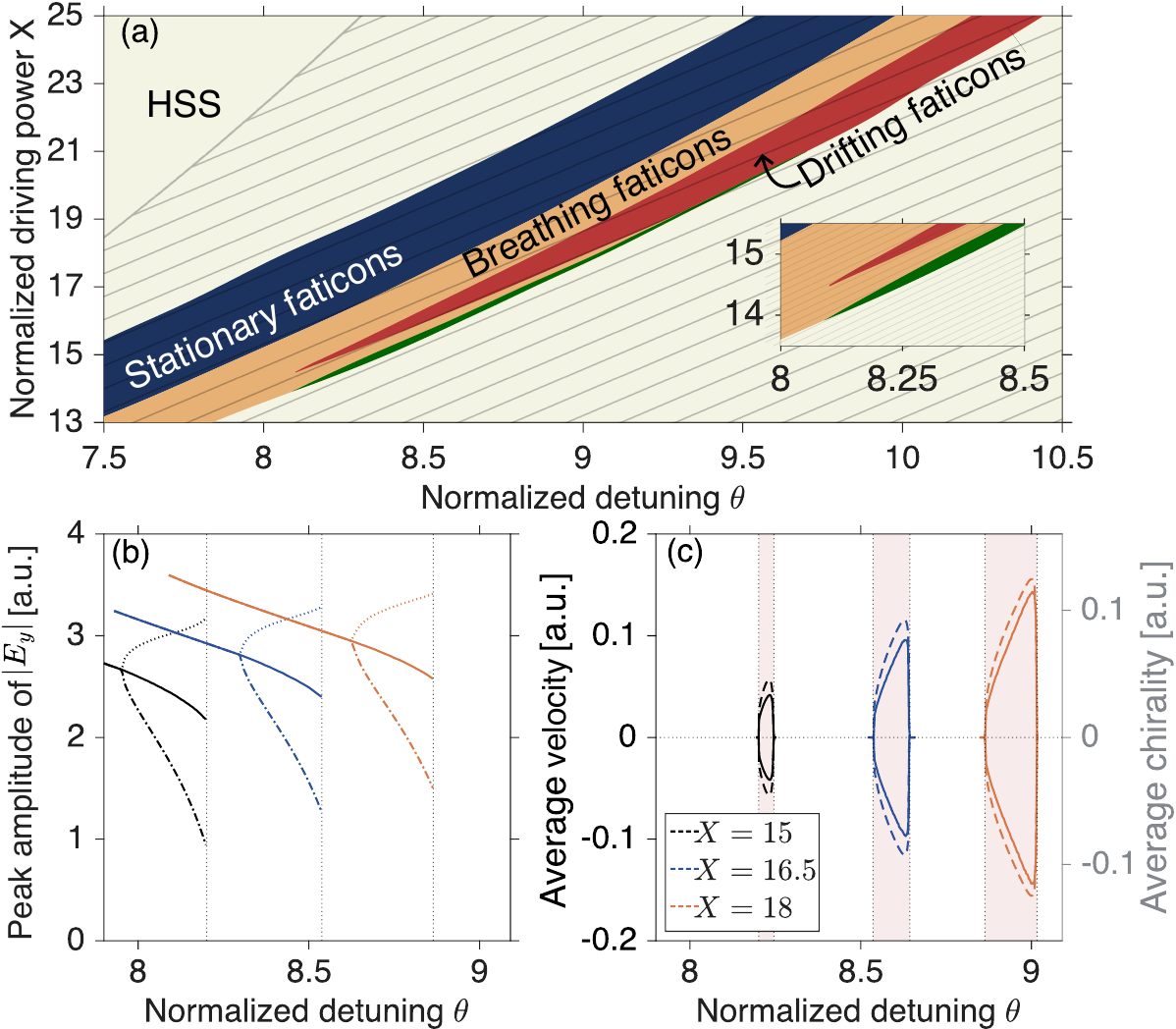}
    \caption{(a) Phase diagram of Eq.~\eqref{eq:LLE} in the parameter space of the cavity detuning $\theta$ and driving power $X$, showing the regions of existence of HSS, stationary faticons (blue region), breathing faticons (yellow region), and drifting breathing faticons (orange region). The dashed region indicates the modulation instability of the symmetry-broken HSS. The inset is a magnification revealing the existence of a green region where period doubling and chaos of the oscillating faticon take place. (b) Peak amplitude $\max_|E_y(t,\tau)|$ of the breathing faticons versus $\theta$ for $X=15$ (black), $X=16.5$ (blue), and $X=18$ (orange), showing the continuous growth of the limit-cycle amplitude beyond the Hopf bifurcation until the NIBT bifurcation point. 
    Dotted, solid, and dashed lines correspond to the maximum, mean, and minimum peak amplitude, respectively.
    (c) Average drift velocity $\bar{v}$ (solid lines) and chirality $\bar{\chi}$ (dashed lines) versus the cavity detuning for the three values of $X$ of (b). 
    }
    \label{fig2}
\end{figure}

This sequence persists over a wide parameter range, as shown by the phase diagram of the long-time dynamics of Eq.~\eqref{eq:LLE} in the $(\theta,X)$ space, Fig.~\ref{fig2}(a).
The existence of polarization faticons is associated with the coexistence of two symmetry-related asymmetric HSS, which provide the two polarization backgrounds connected by the domain walls in the localized structure~\cite{LucasPRL2025}.
The dashed area in Fig.~\ref{fig2}(a) indicates the region in which these asymmetric HSS solutions are modulationally unstable.

Five asymptotic regimes are identified in Fig.~\ref{fig2}: HSS, stationary faticons with a trapped Ising wall (blue region), fixed-position oscillating faticons with a trapped Ising wall (orange region), drifting and breathing faticons with a trapped Bloch wall (red region), and period doubled/chaotic faticons (green region in the magnification in Fig.~\ref{fig2}(a)).
A stationary Ising-type faticon first becomes an oscillating Ising state through a Hopf bifurcation and subsequently acquires a finite propagation velocity through a parity-breaking instability, which transforms the oscillating Ising wall into an oscillating Bloch wall.
Starting from the stationary-faticon region, an increasing detuning first destabilizes the localized Ising state through a supercritical Hopf bifurcation. The resulting faticon oscillates periodically while its position and singularity remain fixed, i.e. an oscillating Ising wall inside a bright soliton. 
This is shown in Fig.~\ref{fig2}(b) where the maximum and minimum amplitudes, as well as the averages, of the temporal oscillations of $|E_y|$ are plotted as dotted, dashed and solid lines for three values of the drive $X$, respectively. 

The onset of translational motion at the NIBT is quantified in Fig.~\ref{fig2}(c), where the average faticon velocity $\bar{v}$ is presented as a function of $\theta$ for three values of $X$.
The average velocity was obtained from the long-time displacement of the wall position after discarding initial transients. 
In the slow-time dynamics of this regime, the two peaks of the $E_y$ field oscillate in antiphase with one of them displaying an average maximum amplitude larger than the other.
This results in a translational velocity that is periodically smaller and higher than the average value corresponding to speeds that first increase and then decrease.
In the stationary and fixed-position breathing regimes, $\bar{v}$ is equal to zero consistent with the reflection symmetry of the Ising localized state. 
Beyond a critical detuning, two branches with opposite average velocities emerge thus providing direct evidence of spontaneous $\mathcal{Z}$ symmetry breaking.
The velocity therefore acts as an order parameter for the drift transition, as expected for an Ising–Bloch bifurcation~\cite{Meron1998,Coullet2002, Clerc2005,deValcarcel2002}.

To characterize the associated transformation of the domain-wall core, we introduce the chirality~\cite{PrezArjona2004},
\begin{equation}
\chi(t)= \text{Im}\left.\left(E_y^*(\tau,t)\partial_\tau E_y(\tau,t)\right)
\right|_{\tau=\tau_c(t)}
\end{equation}
where $\tau_c(t)$ is the instantaneous position of the wall center, identified from the local minimum of $|E_y|$. 
As the cavity soliton breathes, its handedness is characterized through the long-time average. 
Figure~\ref{fig2}(c) shows the average chirality $\bar{\chi}$ as a function of detuning overlaid on the average velocity.
The stationary and fixed-position breathing faticons have zero average chirality, consistent with their reflection-symmetric Ising-type core.
At the NIBT, two branches with opposite chirality emerge together with the positive- and negative-velocity branches. 
The correspondence between their signs demonstrates that the direction of propagation is selected by the handedness acquired by the polarization domain wall.

\begin{figure}
    \centering
    \includegraphics[width=\linewidth]{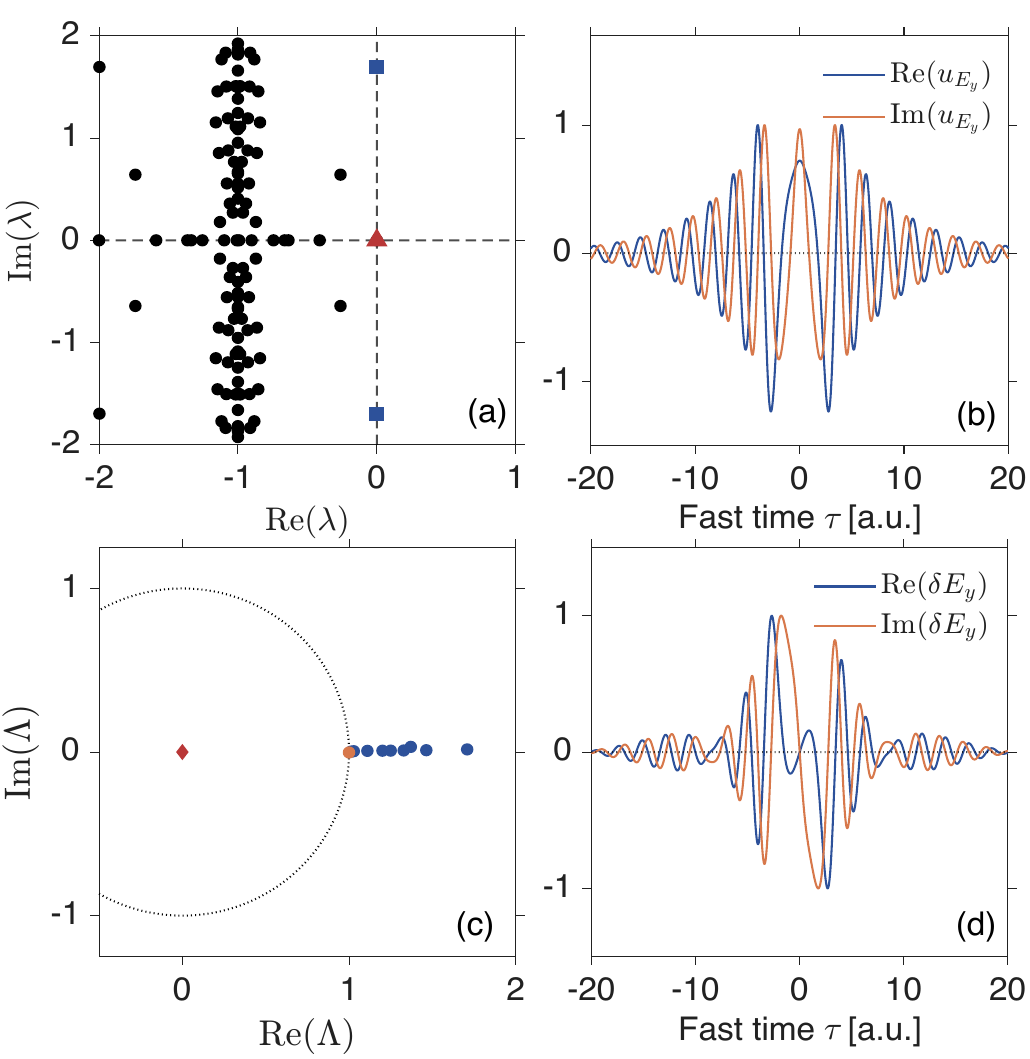}
    \caption{Spectral mechanisms underlying the Hopf and Ising–Bloch bifurcations. 
    (a) Eigenvalue spectrum of the Jacobian evaluated along the stationary-faticon branch immediately below the Hopf bifurcation ($\theta_\mathrm{Hopf}= 8.297$). 
    The eigenvalue at the origin (red triangle) corresponds to the translational Goldstone mode, while the complex-conjugate pair (blue squares) with a small positive real part is responsible for the Hopf instability. 
    (b) Real and imaginary parts of the $E_y$ component of the Goldstone eigenvector $u$.
    (c) Complex-plane trajectory of the relevant (unstable) Floquet multiplier $\Lambda$ of the finite-time linear propagator, evaluated at times multiples of the breathing period $T$.
    Amplification is indicated by a modulus larger than unity after the NIBT ($\theta \approx 8.538$), followed by an approach towards the unit circle.
    (d) Real and imaginary parts of the $E_y$ component of the associated Floquet eigenvector as functions of the fast time $\tau$. The results were computed for $X = 16.5$.}
    \label{fig3}
\end{figure}

The stability spectra that determine the modes responsible for the Hopf bifurcation and the NIBT are shown in Fig.~\ref{fig3}. 
To confirm the Hopf bifurcation at \hbox{$\theta = \theta_\mathrm{Hopf} = 8.297$ for $X=16.5$}, we initialize a stationary breathing faticon and introduce a small perturbation around the localized solution, $E_y^{(0)}(\tau)$, according to \hbox{$ E_y(\tau,t) = E_y^{(0)}(\tau) + \delta E_y(\tau,t)$}. 
At linear order, its evolution is governed by \hbox{$\partial_t \delta E_y = J(E_y^{(0)})\delta E_y$} where $J$ is the Jacobian operator evaluated at the stationary solution. 
Figure~\ref{fig3}(a) shows the spectrum of $J$ immediately below the Hopf bifurcation ($\theta  < \theta_\mathrm{Hopf}$). 
The spectrum contains an eigenvalue at the origin (red triangle) and a complex-conjugate pair of eigenvalues located close to the imaginary axis (blue squares). 
The latter crosses into the right half of the complex plane at the Hopf threshold, destabilizing the stationary faticon and generating the breathing state observed in Figs.~\ref{fig1}(d) and \ref{fig2}(b). 
Their symmetries are analyzed in the Sup. Mat.~\cite{SI}.
The zero eigenvalue originates from the continuous translational invariance of the coupled LLEs and corresponds to the Goldstone mode. 
If $E_y^{(0)}(\tau)$ is a stationary solution, then $E_y^{(0)}(\tau+\delta\tau)$ is also a solution for any spatial displacement $\delta\tau$.
Figure~\ref{fig3}(b) shows the real and imaginary parts of the $y$-polarized component of the Goldstone eigenvector immediately below the Hopf bifurcation, confirming that the neutral perturbation corresponds to an infinitesimal displacement of the entire polarization faticon rather than to an internal deformation of its domain-wall core.

Once the Hopf bifurcation has occurred, the relevant reference state is no longer stationary but periodic in the slow time, i.e., $E_y(\tau,t+T)= E_y(\tau,t)$ where $T$ is the breathing period. Its stability must therefore be analyzed using Floquet theory.
Linear perturbations evolve according to the time-periodic variational equation $\partial_t\delta E_y = J(t)\delta E_y$, where $J(t+T)=J(t)$.
Let $\mathcal{U}(t,0)$ denote the corresponding evolution operator. 
The monodromy operator is defined as $\mathcal{M} = \mathcal{U}(T; t=0)$, with $\mathcal{M}\boldsymbol{\phi}_j=\Lambda_j\boldsymbol{\phi}_j$, where $\Lambda_j$ and $\boldsymbol{\phi}_j$ are the Floquet multipliers and modes, respectively~\cite{DaCunha2011}.
The periodic solution is linearly stable when all non-neutral Floquet multipliers satisfy $|\Lambda_j|<1$.
Neutral multipliers at $\Lambda=1$ arise from the arbitrary temporal phase of the periodic orbit and from its arbitrary position along the $\tau$ coordinate as the system is translationally invariant.

Figure~\ref{fig3}(c) shows the ten largest Floquet multipliers of the evolution operator $\mathcal{U}(t,0)$ for different multiples of the breathing period $T$, just after the NIBT ($\theta\approx8.348$).
The modulus of the largest multiplier exceeds unity (see the evolution of the blue circles at different periods from $1.7$ and then moving towards one) indicating amplification of the perturbation, while the next nine multipliers remain anchored around zero at all periods (see red diamonds).
The components of $E_y$ of the unstable Floquet eigenvector are shown in Fig.~\ref{fig3}(d).
The unstable eigenvector (Fig.~\ref{fig3}(d)) drives the faticon toward a drifting, symmetry-broken state, whose leading multiplier in the co-moving frame (End Matter) is the neutral Goldstone multiplier $\Lambda=1$ (orange circle in Fig.~\ref{fig3}(c)).
The crossing of the unit circle of the largest Floquet multiplier produces two symmetry-related drifting solutions with opposite velocities and chiralities (oscillating Bloch walls) as shown in Fig.~\ref{fig2}(c).

For $14\lesssim X\lesssim 19$, the drifting region is bounded at larger $\theta$ by a narrow band (green in the inset of Fig.~\ref{fig2}(a)) where a period-doubling bifurcation halves the breathing frequency and initiates a cascade~\cite{SI} to irregular spatio-temporal oscillations, as shown in Fig. \ref{fig1}(f).
In this region, the faticon alternates between positive and negative velocities, so that $\bar v=0$. 
At larger $\theta$, localized states cease to exist and the system converges to HSS.

\emph{Conclusions}--- We demonstrated a sequence of dynamical bifurcations experienced by polarization faticons in a continuously driven Kerr resonator. 
Starting from a stationary CS with a trapped Ising domain wall in the $E_y$ field component, increasing the cavity detuning first induces a Hopf bifurcation and generates a spatially stationary breathing faticon.
A second symmetry-breaking instability produces oscillating Bloch walls that are drifting with opposite velocities and domain-wall chiralities.
Linear stability analysis identifies a complex-conjugate eigenvalue pair responsible for the Hopf bifurcation and a translationally neutral eigenvalue associated with the Goldstone mode of the localized state.
Floquet analysis of the breathing faticon links the onset of drift to a parity-breaking mode whose multiplier leaves the unit circle, identifying the instability as a dynamical NIBT of a periodic localized structure. 
This extends the Ising-Bloch scenario from conventional stationary walls to breathing walls trapped in birght CSs, and shows that polarization symmetry breaking can even generate localized spatio-temporal chaos in coherently driven Kerr resonators.
As the onset, speed, and internal dynamics of the
drift are set by controllable detunings and driving powers, these effects are experimentally accessible in two-polarization ring resonators, including those with a phase defect that converts one circular polarization into the other at each round trip~\cite{LucasPRL2025,Coen2024}, where the same behavior of the NIBT persists (see~\cite{SI}).

\emph{Acknowledgments}---
We thank J. Fatome, S. Coen, and J. Gohsrich for useful discussions. 
We acknowledge support from the MPG Lise Meitner Excellence Program 2.0. J.D.M.-V. and F.K.K. are part of the Max Planck School of Photonics supported by the Dieter Schwarz Foundation, the German Federal Ministry of Research, Technology and Space (BMFTR), and the Max Planck Society. L.H. and F.K.K. also acknowledge funding from the European Union's ERC Starting Grant “NTopQuant” (101116680). 

\bibliography{References}

\newpage 

\section{End Matter}
\begin{figure}[h]
    \centering
    \includegraphics[width=\linewidth]{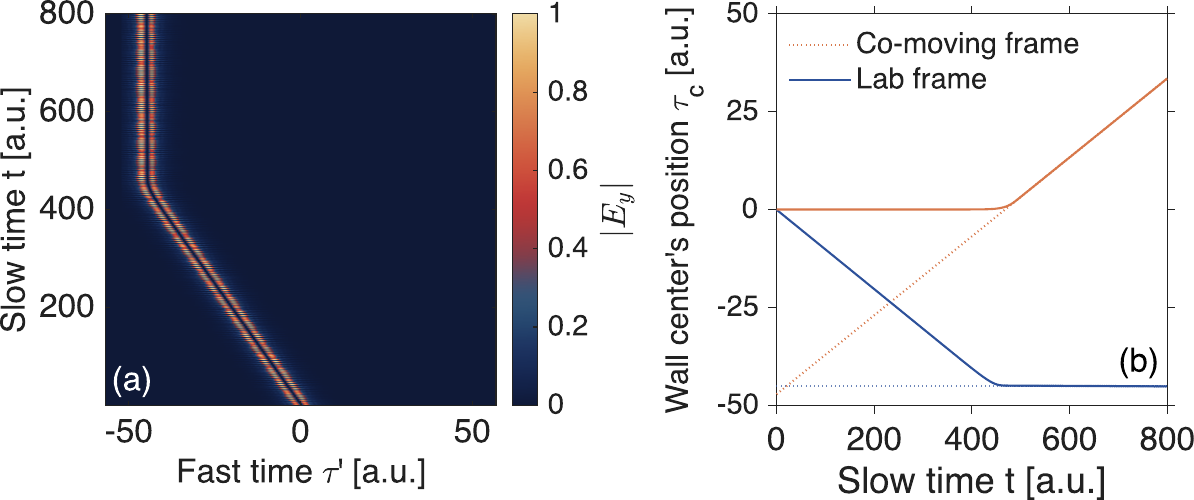}
    \caption{(a) Time evolution of $E_y$ in a co-moving reference frame with speed $v$ after the IBT.
    (b) Position $\tau_c$ of the minimum of $E_y$ as function of the slow time for the faticon in the lab (blue line) and co-moving (orange line) frames. The calculations were computed for $X = 16.5$ and $\theta=8.6$.}
    \label{fig:comparison}
\end{figure}

To characterize the motion of faticons after the non-equilibrium IBT, we compare its evolution in the laboratory and co-moving reference frames.
In Fig.~\ref{fig1}(e) we showed the slow-time evolution of $|E_y(\tau,t)|$ in the laboratory frame for $B=2$, $X=16.5$, and $\theta=8.6$. The wall's center position, denoted by $\tau_c(t)$, is determined from the position of the local minimum of $|E_y(\tau,t)|$.
After a transient, the wall position is accurately described by $\tau_c(t)=v t+\tau_0$, where $v$ is the average propagation velocity and $\tau_0$ is a constant offset. 
In a co-moving frame, Eq.~(\ref{eq:LLE}) becomes 
\begin{eqnarray}\label{eq:LLE2}
\frac{\partial E_{\pm}}{\partial t} &=&  \Big[-\ii\eta\frac{\partial^2}{\partial \tau'^2} + \ii\left(|E_{\pm}|^2 + B|E_{\mp}|^2 \right)-1-\ii\theta\Big]E_{\pm} \nonumber \\
&+& v\frac{\partial E_{\pm}}{\partial \tau'} + \sqrt{X/2} 
\end{eqnarray}
where $\tau'(t,\tau) = \tau - vt$. The corresponding evolution in the co-moving frame is shown in Fig.~\ref{fig:comparison}(a).
In this frame, after the transient time, the position of the wall remains fixed, while the field profile continues to oscillate periodically.
Figure~\ref{fig:comparison}(b) compares the wall center's position $\tau_c$ for the profiles in both reference frames as functions of the slow time $t$.

\newpage
\onecolumngrid

\clearpage

\setcounter{figure}{0}
\renewcommand{\thefigure}{S\arabic{figure}}
\renewcommand{\theequation}{S\arabic{equation}}
\setcounter{equation}{0}
\setcounter{section}{0}
\setcounter{table}{0}

\allowdisplaybreaks

\part*{\large\centering Supplementary Material:\\ Non-Equilibrium Ising-Bloch Transition in Polarization Cavity Solitons}
\begin{center}
    Juan Diego Mazo-Vásquez, Lewis Hill, Flore K. Kunst, and Gian-Luca Oppo
\end{center}

\section{Polarization domain walls for purely symmetric LLEs}

\begin{figure}[hbt!]
    \centering
    \includegraphics[width=0.8\linewidth]{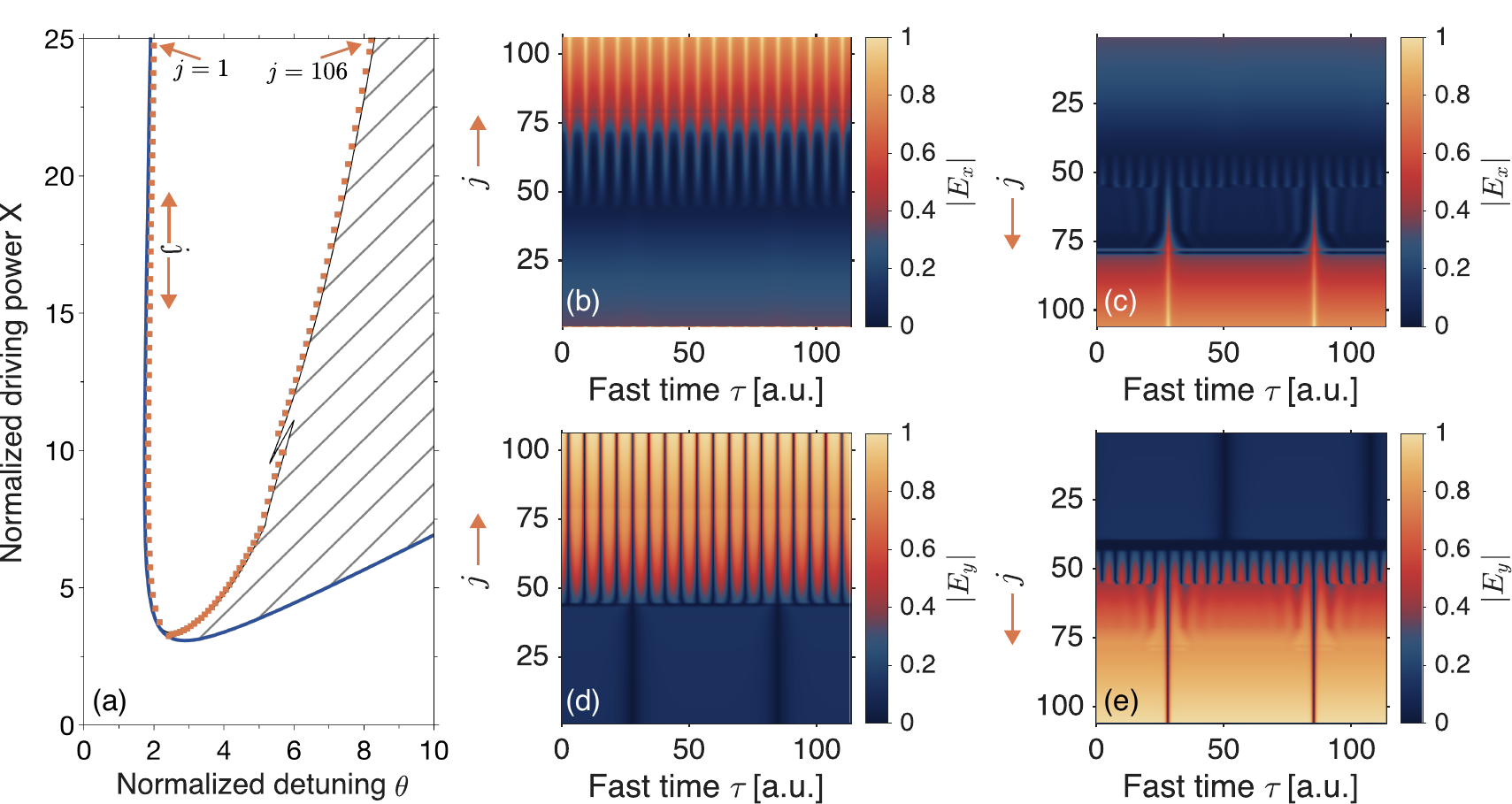}
    \caption{Occurrence of Ising-type polarization domain walls under parameter scans. (a) Parameter space spanned by the detuning $\theta$ and the driving power $X$. 
    The solid blue line corresponds to the spontaneous symmetry bifurcation line, and the dashed area corresponds to the modulation instability region of the asymmetric homogeneous stationary-state solutions to the coupled Lugiato-Lefever equations (Eq.~(1) in the main text). 
    The orange dots correspond to the points of the scan. The counter $j$ indicates the direction of the scan: $j = 0$ and $ j = 106$ correspond to the uppermost left and right points, respectively. 
    Panels (b) and (c) ((d) and (e)) show the magnitude of the $x$-component ($y$-component) of the field across the scan: (b) and (d) increasing $j$, whereas (c) and (e) for decreasing $j$.}
    \label{fig:parameterScan}
\end{figure}

In the main text, we state that stationary polarization domain walls (PDWs) have been broadly studied in Kerr-ring resonators in which orthogonally-polarized fields co-propagate~\cite{GarbinPRL2021, Fatome2023, Wang2026}. In these papers, no spontaneously traveling PDWs have been reported. Here we perform a parameter scan around the spontaneous symmetry bifurcation line and the modulation instability region in the parameter space, as shown in Fig.~\ref{fig:parameterScan} to verify that in the parameter space investigated in the main text, there are no moving PDWs. Across the scanned region (indicated by orange dots in Fig.~\ref{fig:parameterScan}(a)), only Ising-like stationary PDWs were found. 
When the scan starts close to the SBS bifurcation line (from the upper left point in Fig.~\ref{fig:parameterScan} (a)), a flat solution for the $E_x$ component, and a two-PDW profile for the $E_y$ component evolve into stationary patterns via a modulation instability, as shown in panels (b) and (d). 
In contrast, when the scan starts close to the modulation instability region (from the upper right point in Fig.~\ref{fig:parameterScan} (a)), both linearly
polarized components $E_x$ and $E_y$ initially maintain the profiles of two stable PDWs during the scan. Later, the linearly polarized component $E_x$ evolves into a flat profile whereas $E_y$ preserves a two PDWs profile after undergoing a transformation into a stationary pattern. No moving PDWs were found in these scans.

\section{Linear stability analysis and symmetry of the Hopf Eigenvectors}
Here we perform linear stability analyses on generic stationary solutions of the coupled LLEs, Eq.~(1) in the main text. We first write the differential equations in terms of the linearly polarized $x$ and $y$ components, i.e., \hbox{$E_x = (E_+ + E_-)/\sqrt{2}$} and \hbox{$E_y = (E_+-E_-)/(\ii\sqrt{2})$} leading to 
\begin{equation}\label{LLE-xy}
    \begin{split}
    \frac{\partial E_x}{\partial t} &= \sqrt{X}-(1+\ii \theta) E_x - \ii \eta \frac{\partial^2}{\partial \tau^2}+\frac{\ii}{2}\left[(B+1)|E_x|^2 E_x + 2 |E_y|^2 E_x + (B-1) E_y^2 E_x^*\right],\\
    \frac{\partial E_y}{\partial t} &= -(1+\ii \theta) E_y - \ii \eta \frac{\partial^2}{\partial \tau^2}+\frac{\ii}{2}\left[(B+1)|E_y|^2 E_y + 2 |E_x|^2E_y + (B-1)E_x^2 E_y^*\right].
    \end{split}
\end{equation}
The field components are then separated in their corresponding real and imaginary parts, namely $E_{x,y} = a_{x,y} + \ii b_{x,y}$, where $a_{x,y} = \mathrm{Re}(E_{x,y})$, and $b_{x,y} = \mathrm{Im}(E_{x,y})$. 

We write the perturbations to the stationary states of Eq.~\eqref{LLE-xy} as $a_{x,y} = a_{x,y}^{(0)} + \delta a_{x,y}$, and $b_{x,y} = b_{x,y}^{(0)} + \delta b_{x,y}$ where $a_{x,y}^{(0)}$ and $b_{x,y}^{(0)}$  are the stationary-state solutions for $a_{x,y}$ and $b_{x,y}$, respectively.
To assess the stability of the stationary states, we transform the linearized equations from the fast-time domain $\tau$ to the Fourier domain, where $\partial_\tau^2 \rightarrow - k^2$, with $k$ being the associated wave number of the variable $\tau$. The time evolution of the infinitesimal perturbations $\vec\delta$ is given by
\begin{equation}
    \frac{\partial \vec{\delta}}{\partial t} = J(k^2) \; \vec{\delta},
\end{equation}
where $\vec{\delta} = \left(\delta a_x\, , \delta b_x\, , \delta a_y\,, \delta b_y\right)^T$, and $J(k^2)$ is the Jacobian matrix of the system that depends on the squared magnitude of the Fourier wavevector. In the fast-time domain, we consider $\vec{q}_n$, the $n$-th eigenvector associated to the eigenvalue $\lambda_n$ of the Jacobian $J$ and defined as 
\begin{equation}
    \vec{q}_n = \left(a_x(\tau)\, , \, b_x(\tau)\, , \, a_y(\tau)\, , \, b_y(\tau) \right)^T,
\end{equation}
as well as $\mathcal{P}$ and $\mathcal{Z}$ the parity and permutation operators defined as $\mathcal{P}: \tau \rightarrow -\tau$, and \hbox{$\mathcal{Z}: (E_x, E_y) \rightarrow (E_x, -E_y)$}, respectively.
Under the transformation $\mathcal{Z}\circ\mathcal{P}$, the vector $q_n$ becomes
\begin{equation}
    \mathcal{Z}\circ\mathcal{P}\, (\vec{q}_n) =  \left(a_x(-\tau)\, , \, b_x(-\tau)\, , \, -a_y (-\tau)\, , \, -b_y(-\tau)\right)^T.
\end{equation}
For the eigenvectors associated to the Hopf bifurcation, whose eigenvalues are indicated by a blue squares in Fig.~3 (a) of the main text, the eigenvalue of the operator $\mathcal{P}\circ\mathcal{Z}$ is $+1$, therefore indicating that those states are symmetric under $\mathcal{Z}\circ\mathcal{P}$, a characteristic of Ising fronts~\cite{MichaelisPRE2001}. 
This feature allows us to define the oscillating-Ising states.

\begin{figure}[hbt!]
    \centering
    \includegraphics[width=0.55\linewidth]{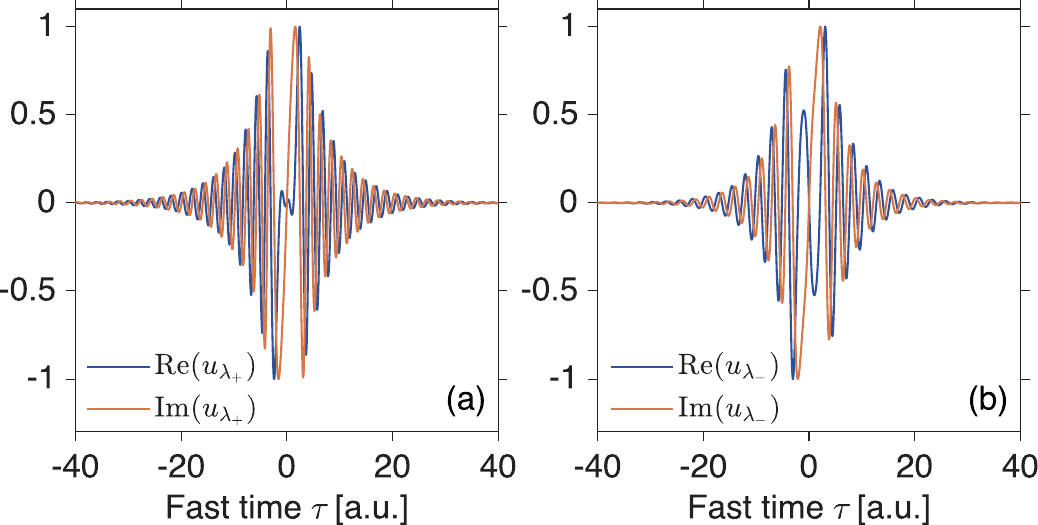}
    \caption{Real and imaginary parts of the eigenvectors associated to the $y$-linearly polarized component of the field, corresponding to the eigenvalues $\lambda_\pm$ marked in Fig.~3(a) of the main text with blue squares.}
    \label{fig:evecs}
\end{figure}

Figure~\ref{fig:evecs} shows the real and imaginary parts of the eigenvectors associated to $E_y$ corresponding to the eigenvalues which cross into the real part of the complex plane in the Hopf bifurcation, marked in Fig.~3(a) of the main text with a blue square, and an orange diamond.

\section{Regime of chaotic dynamics}

\begin{figure}[hbt!]
    \centering
    \includegraphics[width=\linewidth]{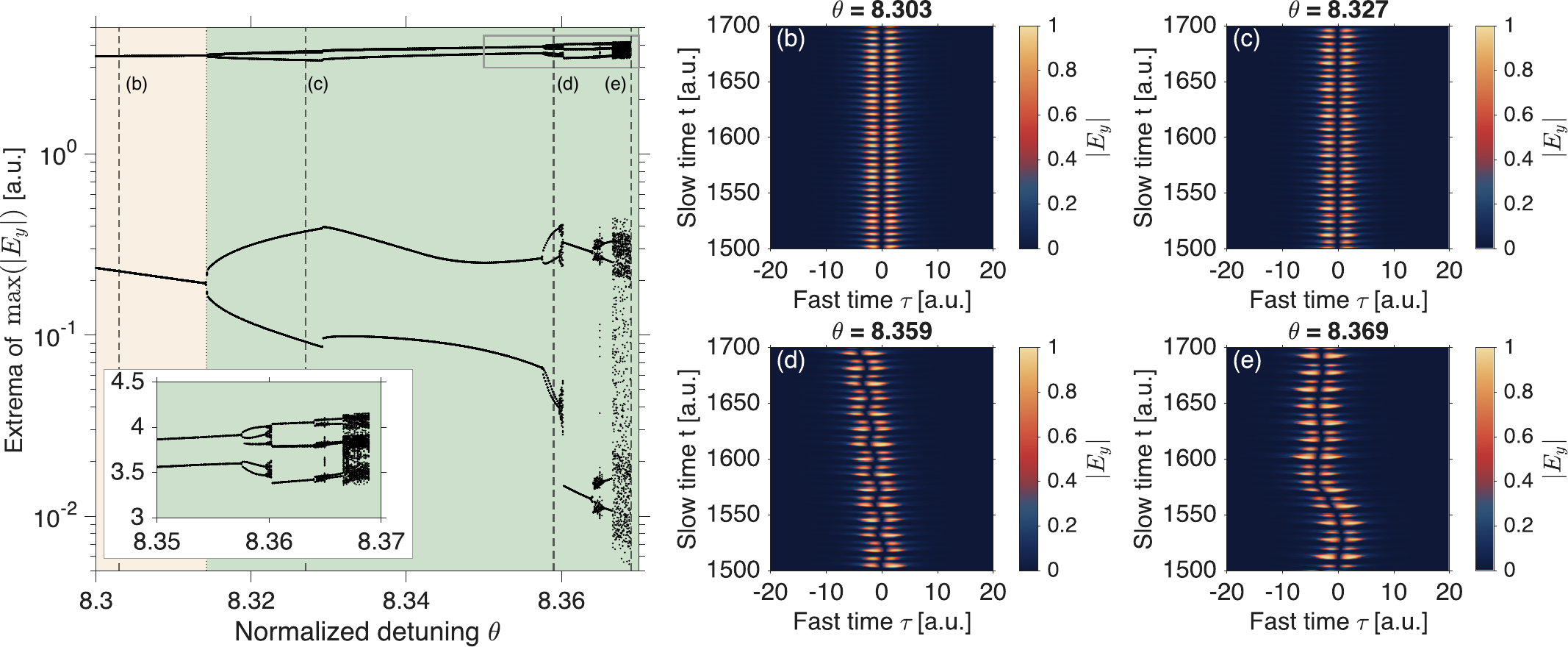}
    \caption{Extrema of the maxima of the field amplitude $|E_y|$ for a cavity detuning scan, at $X = 15$. The orange and green shaded areas correspond to the breathing faticon and \textit{jittering faticon} regions in Fig.~2 of the main text. 
    The vertical dashed lines indicate the values of $\theta$ used to compute the results shown in panels (b)-(e). 
    The inset corresponds to a zoom into the region indicated by the rectangle in the upper-right corner. }
    \label{fig:chaotic}
\end{figure}

Fig.~2 in the main text shows different behaviour of the solutions of Eq. (1) in the parameter space spanned by the cavity detuning $\theta$ and driving power $X$. The green area in this parameter space denoted as {\it jittering faticon region} corresponds to period-doubled and spatio-temporally chaotic faticons. 
Here we show that in this region a cascade of period doubling bifurcations occurs leading to spatio-temporal chaotic dynamics. 
Figure~\ref{fig:chaotic}(a) shows the extrema of the maxima of the magnitude of the field component $E_y$ for a cavity detuning scan at $X=15$.
For $\theta < 8.314$, the faticon is breathing but remains stationary in the Ising phase, as shown in panel (b). 
At $\theta \approx 8.314$ a period doubling bifurcation occurs, but the faticon remains in the Ising phase (panel (c)).
At around $\theta \approx 8.357$, a second period doubling bifurcation occurs, followed by a cascade of period doublings and chaos at $\theta \approx 8.359$. A window of periodic oscillations is reached at around $\theta \approx 8.360$. 
In this regime, the faticon undergoes complex spatial oscillations while drifting, as shown in panel (d).
Regular dynamics remains up to $\theta \approx 8.363$, from which another period doubling cascade leads to spatio-temporal chaotic dynamics, as shown in panel (e). 

\section{Simulations including a phase defect}
\begin{figure}[hbt!]
    \centering
   \includegraphics[width=0.7\linewidth]{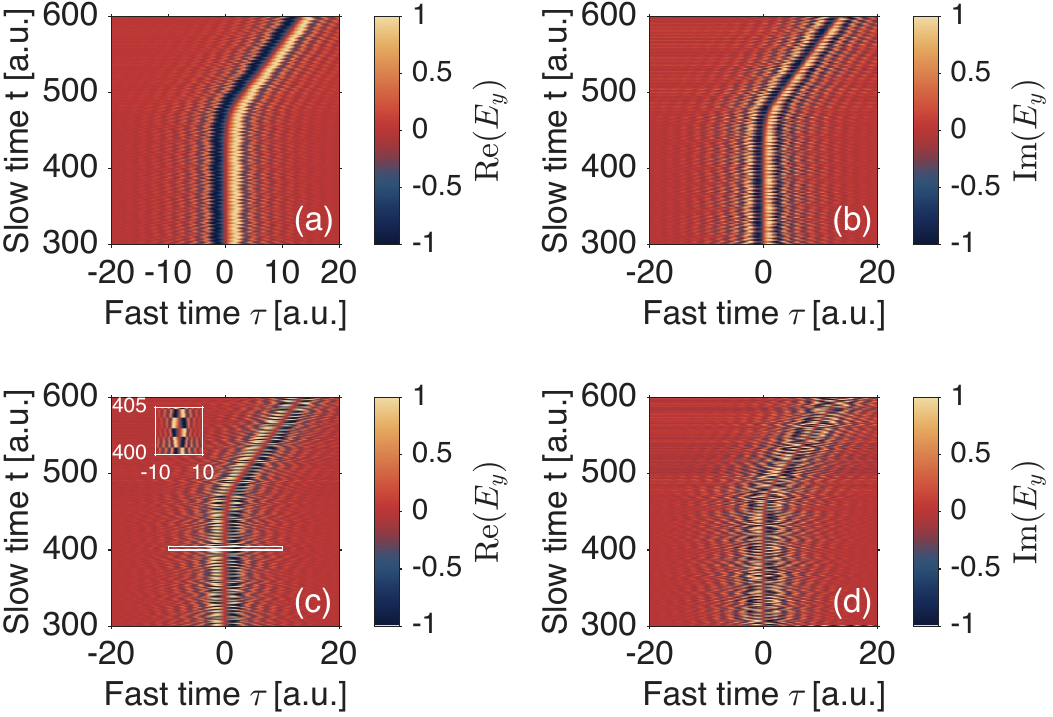}
    \caption{Time evolution of the real and imaginary parts of the linearly polarized component $E_y$ for $\theta = 8.6$, and $X = 16.5$. Panels (a) and (b) ((c) and (d)) without (with) a phase defect. The inset in panel (c) corresponds a zoomed-in view of the region $\tau \in [-10,10]$ for $t \in [400,405]$, indicated by the rectangle. }
    \label{fig:impurity}
\end{figure}

Here we present the results of time evolutions of the faticons after the NIBT when there is a phase defect in the resonator, a configuration discussed in Refs.~\cite{Coen2024, LucasPRL2025}. 
The phase defect corresponds to a polarization controller implementing a $\pi$ phase-shift and causing the circularly polarized fields $E_\pm$ to swap at each roundtrip. To simulate this defect, in our numerical procedure we swap the fields $E_+ $ and $ E_-$ every cavity period $t = 1$ in the slow time.
Figure~\ref{fig:impurity} shows the time evolution of the real and imaginary parts of the $E_y$ component of a field circulating inside of a ring resonator, without adefect (panels (a) and (b)), and with a defect (panels (c) and (d)). 
The results show that the motion of the faticon after the NIBT persists regardless of the phase defect, even the time of the onset of the drift does not change with respect to the case in which the defect is absent. 

\end{document}